\documentstyle[12pt]{article}
\begin{document}
\begin{center}
{\large\bf  Low-Temperature Quantum Critical Behaviour}\\
{\large\bf  of Systems with Transverse Ising-like}\\
{\large\bf   Intrinsic Dynamics.}\\
{\baselineskip 24pt} \vspace{1cm} A. Caramico D'Auria$^a$, L. De
Cesare$^b$, I. Rabuffo$^b$ \vspace{0.5cm}
\end{center}
{\it $^a$ Dipartimento di Scienze Fisiche,Universit\`a di Napoli,
80125 Napoli, Italy and Istituto Nazionale
per la Fisica della Materia, Unit\`a di Napoli, Italy \\
\it $^b$ Dipartimento di Fisica ``E.R. Caianiello", Universit\`a di
Salerno, Salerno, Italy and Istituto Nazionale per la Fisica
della Materia, Unit\`a di Salerno, Italy\\} \vspace{1cm}
\vspace{0.2cm}

Corresponding author: I. Rabuffo, Dipartimento di Fisica ``E.R.
Caianiello", via S. Allende 84081, Baronissi (Italy). Phone +39-089965392,
Fax +39-089965275, e-mail Rabuffo@sa.infn.it

\vspace{0.5cm}
PACS NUMBERS: 68.35.Rh;05.10.Cc;75.10.Jm
\vspace{0.3cm}


\vspace{1cm}

\centerline{\Large\bf Abstract} The low-temperature properties and
crossover phenomena of $d$-dimensional transverse Ising-like
systems within the influence domain of the quantum critical point
are investigated solving the appropriate one-loop renormalization
group equations. The phase diagram is obtained near and at $d=3$
and several sets of critical exponents are determined which
describe different responses of a system to quantum fluctuations
according to the way of approaching the quantum critical point.
The results are in remarkable agreement with experiments for a
wide variety of compounds exhibiting a quantum phase transition,
as the ferroelectric oxides and other displacive systems.

\newpage
\baselineskip 24pt

\section{Introduction}

\hspace {16pt} The topic of quantum phase transitions (QPT's),
where the order is destroyed solely by quantum fluctuations, is a
very active field in condensed matter physics research [1]. These
transitions, which occur only at temperature $T=0$ varying a
nonthermal control parameter, are considered to play an important
role for understanding the unconventional behaviour of many
quantum systems at low temperatures such as non-trivial power laws
or non-Fermi liquid physics. Indeed, the existence of an influence
domain around a $T=0$ quantum critical point (QCP) is thought to
be relevant for gaining insight into several low temperature
phenomena using the reliable foundations of the modern theory of
critical phenomena.

Recent theoretical studies [1-7] have shown that a lot of
information can be extracted from the temperature-dependent
Renormalization Group (RG) equations as an expression of the most
reliable theory of thermal and quantum critical fluctuations.
Along this direction, a rich low-temperature scenario was obtained
in Refs. [4-6]  for the
 case of an interacting Bose gas with the chemical potential
fixed at its QCP value.

Unfortunately, the inherent simplicity of the RG equations for
bosonic systems is lost when one considers other microscopic
models exhibiting a QPT. Important examples are the transverse
Ising model (TIM)[8] and the quantum displacive models [9,10].
These models and related ones [5,8,11] can be described in an
unified way also near the QCP and the RG treatment offers the
possibility to obtain the low-temperature properties in the domain
of influence of the QCP near and at $d=3$ dimensions taking
correctly into account the competing quantum and thermal
fluctuations. Nevertheless, the problem does not appear so simple
to be solved as for bosonic systems and some attention must be
paid to extract the correct low-temperature quantum critical
behaviour of the relevant thermodynamic quantities.

The aim of the present paper  is to
give a solution of the problem near and at $d=3$ by  solving the
appropriate RG equations in the low temperature limit. This  allows us to obtain
asimpotically exact results so that a suitable comparison with the
experiments becomes possible.

The outline of this paper is as follows. In section 2 we introduce
the model together with the $T$-dependent one-loop RG equations
and give a summary of the $(T=0)$-quantum critical predictions
relevant for our purposes. Section 3 is devoted to derive the
solution of the RG equations in the low-temperature regime to
leading order in the interaction  parameter. In section 4 the
global expressions of the relevant thermodynamic quantities and
the phase diagram of the models in the influence domain of the QCP
are obtained. The low-temperature quantum critical properties are
considered in section 5 and the related crossover phenomena are
studied in section 6, where a comparison with experiments is also
performed. Finally, in section 7, some concluding remarks are
made.

\section{One-loop renormalization group equations for transverse Ising-like models.}
\setcounter{equation}{0}
\hspace {16pt} We focus our attention on
the  transverse Ising-like models described by the quantum action
(in convenient units):
\begin{equation}
{\cal H}\{ \vec\psi \}= {1\over2} \int d^{d}x \int_{0}^{1/
T_{0}}d\tau \left\{ \left( \nabla\vec\psi(\vec x,\tau)\right)^{2}
+ \left( {\partial \vec\psi ({\vec x},\tau) \over \partial
\tau}\right)^{2}+ r_{o}\vec\psi^{2}({\vec x},\tau )+{u_{o}\over
8}\vec\psi^{4}({\vec x},\tau)\right\}
\end{equation}            %      (2.1)
where $\vec\psi(\vec x,\tau)\equiv \left\{\psi_j(\vec x,\tau);
j=1,\ldots, n\right\}$ is a $n$-vector ordering field and the
coupling parameters $r_0, u_0$ depend on the quantum model under
study: for the TIM and related real
systems  see, for instance, Refs. [8,12]; for quantum displacive systems [9,10]  $r_0$ has to be identified as the
opposite of the strength of interaction parameter $S$ and $u_0>0$
measures the anharmonic part of the interaction.

Now, we can apply the Wilson RG procedure to the action (2.1) in
the $(\vec k, \omega_l)$- Fourier space with $0<|\vec k|<1$.
Reducing  the degrees of the freedom and rescaling  the wave
vectors $\vec k$, the Matsubara frequencies $\omega_l=2\pi lT\
(l=0\pm1, \pm 2, \ldots)$ and the ordering field $\vec\psi (\vec
k,\omega_{l})$, we obtain, to one-loop approximation, the
following T-dependent RG differential equations for the
renormalized parameters $r(l)$, $u(l)$, $T(l)$ (see for instance
Ref.[4]):
 \begin{equation}
   \left\{
   \begin{array}{l}
   \displaystyle {dr(l)\over dl}= 2r(l)+{n+2\over 4}K_d u(l)F_1
   \left(r(l), T(l)\right) \\
    \displaystyle {du(l)\over dl}=\varepsilon u -{n+8\over 4} K_d u^2(l) F_2\left(r(l),
    T(l)\right)\\
    \displaystyle {dT(l)\over dl}=zT (l)
   \end{array}
   \right.
 \end{equation}            %      (2.2)
with $\eta=0$ for the Fisher correlation exponent and
$K_d=2^{1-d}\pi^{-d/ 2} /\Gamma({d\over 2})$, where $\Gamma(x)$
is the gamma function. In Eqs. (2.2), $\varepsilon=3-d$, $z=1$ is
the dynamical exponent for the quantum model and the functions
$F_i(r,T)$ $(i=1,2)$ are given by
\begin{eqnarray}
F_{1}(r,T)&=&{1\over 2}(1+r)^{-{1\over
2}}\coth\left[{(1+r)^{1\over 2} \over 2T}\right]\\            %      (2.3)
F_{2}(r,T)&=&-{\partial F_{1}(r,T)\over
\partial r}={1\over 4}(1+r)^{-{3\over
2}}\coth\left[{(1+r)^{1\over 2} \over 2T}\right]+\nonumber\\
&& +{1\over 8T}(1+r)^{-1}\sinh^{-2}\left[{(1+r)^{1\over 2} \over
2T}\right].
\end{eqnarray}            %      (2.4)
In order to extract all the possible information about
criticality, crossover phenomena and thermodynamic properties of
the models here considered, one should exactly solve the system of
Eqs. (2.2) with the initial conditions:
\begin{equation}
r(0)=r_0,\ u(0)=u_0\ ,\ T(0)=T_0\equiv T\ .
\end{equation}            %      (2.5)

Note that for systems whose functional representation is generated
by a Hubbard-Stratonovich transformation, one expects both $r_0$
and $u_0$ to depend on temperature as, for example, in the TIM
[8]. However, this $T$-dependence is usually negligible in
comparison to that near criticality, so we do not consider it
explicitly through this paper.

Solving  the RG equations (2.2) with the initial conditions (2.5)
is a very difficult task and, to obtain explicit results, one must
resort to suitable approximations or treatable asymptotic regimes.
In the original pioneering treatments [11-13] the interest was
essentially focused on the so called ``quantum" regime at $T=0$
and on the ``classical" one as $T(l)\to\infty$ by iteration of the
RG transformation. Here, for future utilities we summarize the
main predictions which are relevant for us.

The finite-temperature fixed points (FP's) of Eqs.(2.2), expressed
in terms of the most appropriate coupling parameter $v=u\cdot T$,
govern the critical behaviour in the classical regime, which is
well known in the literature [14-16]. Indeed, in such a case, the
quantum degrees of freedom are ineffective and any quantum system
behaves, in terms of the critical deviation $r_0-r_{0c}(T)$, as a
classical one with the same dimensionality and order parameter
symmetry, in agreement with the universality hypothesis. Quantum
effects become relevant as $T\to 0$. In particular, at $T=0$ and
in terms of the nonthermal deviation parameter $r_0-r_{0c}$, a
critical behaviour different from the classical one occurs.

For a generic $(T=0)$-FP $\quad (r^*,u^*)$, the related
eigenvalues to first order in $\varepsilon$ are [5]
\begin{equation}
\left\{
\begin{array}{l}
\displaystyle \lambda_r=2-{n+2\over 16} K_{d} u^* \\
\displaystyle \lambda_u=\varepsilon -{n+8\over 8} K_{d}u^*
\end{array}
\right.
\end{equation}            %      (2.6)
with $r^*=u^*=0$ for the Gaussian FP and $r^*=-{n+2\over
n+8}\varepsilon$, $u^*={16\over (n+8)K_{d}}\varepsilon $ for the
nongaussian one.

One sees that, since in any case $\lambda_r>0$, the
($T$=0)-critical surface and  the related QCP ($T_c=0, r_{0c}$)
are determined by imposing that the linear scaling field
$t_r=r+bu=t_{r_0}e^{\lambda_r}$, with $b={{(n+2})\over 16}K_{d}$,
is zero. This yields
\begin{equation}
r_{0c}=-bu_0\ .
\end{equation}            %      (2.7)
 Then, in terms of variable $t_{r_0}=r_0-r_{0c}$ driving the QPT,
the critical exponents, which govern the quantum critical
behaviour approaching the QCP along the $(T=0)$-isotherm, can be
easily obtained  around $d=3$. For instance, the correlation
length  and the susceptibility exponents $\nu_r={1/ \lambda_r}$
and $\gamma_r=2\nu_r$ (with $\eta=0$), to first order in
$\varepsilon=3-d$, are given by:
\begin{eqnarray}
\nu_r&=&{1\over 2}+{n+2\over 4(n+8)} \varepsilon\quad , \qquad
\gamma_r=1+{n+2\over 2(n+8)} \varepsilon \quad ,\qquad d<3 \\            %      (2.8)
\nu_r&=&{1\over 2}\qquad , \qquad \gamma_r=1 \qquad ,\qquad d>3
\end{eqnarray}            %      (2.9)
For $d=3$, logarithmic corrections to the mean-field exponents
occur. Notice that for $d>3$, the remaining mean-field exponents
can be obtained, as usual, taking into account that $u_0$ is a
dangerous irrelevant variable [14-16].

The previous RG analysis shows that, as $T\to 0$, a dimensional
crossover $d\to d+1$ occurs with
$X_{quantum}(d)=X_{classical}(d+1)$ for a generic critical
exponent $X$.

An important feature to be noted is that the temperature $T$, as
well as the parameter $r_{0}$, is a relevant scaling field
$(T(l)=Te^l)$. Therefore also a small but finite value of $T$
measures a deviation from the ($T$=0)-critical surface. However,
for extracting all the relevant information about the critical
behaviour around the QCP, the parameter $T$ cannot be treated as
$r_0$ since the functions $F_i(r,T)$ cannot be expanded in power
series of $T$ and hence, as $T\to 0$, it is not possible to follow
the usual linearization scheme of the RG analysis. This
unfortunate feature forces to solve the full $T$-dependent RG Eqs.
(2.2), at least in the low-temperature limit. For overcoming this
intrinsic difficulty of the problem  we introduced in ref. [4] the
``ad hoc" concept of ``temperature-dependent linear scaling
fields". Although this idea allowed us to capture the essential
aspects of the low-$T$ quantum criticality for a wide variety of
quantum systems, it appears as a rather ``nonconventional ansatz"
which needs further supporting justifications. The next sections
are just devoted to give such a support; moreover  further
physical predictions are given solving exactly Eq. (2.2) to order
$O(\varepsilon ,u)$ and in the low-$T$ regime, without additional
artificious assumptions.

\section{Solution of the RG equations in the low-temperature
regime to leading order in the interaction parameter.}
\setcounter{equation}{0}
 \hspace {16pt}
 Our aim is to solve the general
one-loop Eqs. (2.2)  in the low-temperature limit to leading order
in $u_0\ll 1$ (with $u_0 \sim O(\varepsilon)$ if $\varepsilon \neq
0$). In our considerations we  assume $u=O(\varepsilon)\ll 1$ and
work for low-$T$ RG flow correct to the order $O(\varepsilon,
u_0)$.

First, let us consider the simplified equation for $u$ (setting in
the second equation of (2.2) $r=0$ and $T=0$):
\begin{equation}
{du\over dl}=\varepsilon u-{n+8\over 16} K_{d} u^2
\end{equation}            %      (3.1)
with the initial condition $u(0)=u_0$. This can be solved exactly
and we have:
\begin{equation}
u(l)={u_0e^{\varepsilon l}\over Q(l)} = O(\varepsilon, u_{0})
\end{equation}            %      (3.2)
where \begin{equation}
 Q(l)=1+{n+8\over 16}K_{d}{u_0\over \varepsilon}\left(e^{\varepsilon
 l}-1\right).
 \end{equation}            %      (3.3)
 In the limit $\varepsilon=0$, Eq.(3.2) gives:
 \begin{equation}
 u(l)={1\over C_0(l+l_0)}
 \end{equation}            %      (3.4)
 where
\begin{equation}
C_0={n+8\over 16}K_3={n+8\over 32\pi}
\end{equation}            %      (3.5)
and
\begin{equation}
  l_0={1\over C_0 u_0}.
\end{equation}            %      (3.6)

It is immediate to see that
\begin{equation}
Q(l)\to\left\{
\begin{array}{l}
1\qquad , \qquad l\to 0\\
\left.\begin{array}{lcl} \displaystyle {n+8\over 16}K_{d} {u_0\over
\varepsilon}e^{\varepsilon l}\qquad &,&\qquad \varepsilon >0\\
\displaystyle 1+{n+8\over 16}K_{d} {u_0\over |\varepsilon|}\qquad &,&
\qquad \varepsilon<0
\end{array} \right\}, \quad  l\to\infty
\end{array}\right.
\end{equation}         %   (3.7)
and hence, as expected:
\begin{equation}
u(l)\to\left\{\begin{array}{l}
u_0\qquad , \qquad l\to 0\\
u^*=\left\{\begin{array}{lcl} \displaystyle {16\over (n+8) K_{d}}\varepsilon\qquad&,&\qquad \varepsilon >0\\
0 \qquad &,& \qquad \varepsilon<0
\end{array}\qquad \right .,\quad l\to\infty
\end{array}\right. .
\end{equation}         %   (3.8)

The solution (3.2) of the simplified Eq. (3.1)  is not, obviously,
a solution of the full equation for $u(l)$ which involves
contributions arising from $r(l)$ and $T(l)$. We now  prove that
the factor $F_2(r(l),T(l))$, previously neglected in the second of
Eqs. (2.2), contributes to the reduced solution  only to the order
$O(\varepsilon^2, u^2_0)$.

Let us assume a full solution of the form
\begin{equation}
  u(l)=\tilde u(l)f(l)
\end{equation}         %   (3.9)
where $\tilde u(l)=O(\varepsilon,u_0)$ denotes the reduced
solution (3.2) and $f(l)$ has to be determined with the initial
condition $f(0)=1$. From Eq. (2.2) we get for $f(l)$ the equation
\begin{equation}
  {df(l)\over dl} = {n+8\over 16} K_{d} \tilde u(l)
  f(l)\left[1-4f(l)F_2\left(r(l),T(l)\right)\right] .
\end{equation}         %   (3.10)
Eq. (3.10) has the ``Bernoulli equation" form
$dy/dx+P(x) y=Q(x)y^n$ and hence we have for $f(l)$ the standard
solution:
\begin{eqnarray}
f(l) &=& e^{{n+8\over 16}K_{d}\int_0^l dl'\tilde u(l')} \times
\nonumber \\
&& \times \left\{ 1+{n+8\over 4} K_{d} \int_0^l dl'\tilde u(l') F_2
(l') e^{-{n+8\over 16}K_{d}\int_0^{l'} dl''\tilde
u(l'')}\right\}^{-1},
\end{eqnarray}         %   (3.11)
where $F_2(l)\equiv F_2(r(l),T(l))$. On the other hand $\tilde
u(l)=O(\varepsilon,u_0)$ and we are interested to obtain only
results correct to the order  $O(\varepsilon,u_0)$. Then, from the formal
expression (3.11) it follows
\begin{eqnarray}
f(l) &=& 1+{n+8\over 16} K_{d} \int_0^l dl'\tilde u(l')
\left(1-4 F_2 (l')\right)+O(\varepsilon^2,u^2_0)\nonumber \\
&=& 1 + O(\varepsilon,u_0)\ .
\end{eqnarray}         %   (3.12)
This allows us to conclude that $u(l)=\tilde u(l)
(1+O(\varepsilon,u_0))=\tilde u(l)+O(\varepsilon^2,u^2_0)$. So,
$F_2(r,T)$-contributions, and hence $T$-dependent terms, do not
enter $u(l)$ to $O(\varepsilon,u_0)$ order.

The problem to integrate the system (2.2) reduces now to solve
only the equation for $r(l)$ with $u(l)\equiv \tilde u(l)$ and
$T(l)=Te^l$, to the order of interest.

An explicit solution for $r(l)$ with  arbitrary $T$ is, of course, hopeless but
the problem becomes accessible working in the low-temperature
regime. The solution can be formally written
as
\begin{equation}
r(l)=r_0 e^{\Lambda_r(l)}h(l)
\end{equation}         %   (3.13)
with
\begin{eqnarray}
\Lambda_r(l)&=& 2l-{n+2\over 4} K_{d} \int_0^ldl'u(l')
F_2^{(0)}(l')\nonumber \\
&=& 2l+O(\varepsilon,u_0) \ ,
\end{eqnarray}         %   (3.14)

\begin{equation}
h(l)=1+{1\over r_0}{n+2\over 4}K_{d}\int _0^l dl' e
^{-\Lambda_r(l')}u(l') F_1^{(0)} (l')
\end{equation}         %   (3.15)
and $F_i^{(0)}(l')\equiv F_i (0,T(l'))$ $(i=1,2)$.

Since $\exp(-\Lambda_r(l))=e^{-2l}+O(\varepsilon,u_0)$ and
$du/dl=O(\varepsilon^2,u^2_0)$, $h(l)$ with integration by parts
in (3.15) becomes:

\begin{equation}
h(l) = 1+ {1\over r_{0}}{{n+2}\over16}K_{d}\left ( u_{0} -
e^{\Lambda_r(l)} u_{l}\right)  + {1\over r_0}{n+2\over 4} K_{d}
\int_0^l dl'{e^{-2l'}u(l')\over e^{1/T(l')}-1}.
\end{equation} %(3.16)
Eq. (3.16) suggests to express more simply the required solution
by means of the linear combination
\begin{equation}
t_r(l)=r(l)+{{n+2}\over 16} K_{d} u(l)
\end{equation}  %(3.17)
and we have:
\begin{equation}
t_r(l)=e^{\Lambda_r(l)}\left\{t_{r}(0)+{n+2\over
4}Kd\int_0^l dl'{e^{-2l'}u(l')\over e^{1/T(l')}-1}\right\}
\end{equation}         %   (3.18)
with $t_{r}(0)=r_{0}+{n+2\over 16}K_{d}u_{0}=r_{0}-r_{0c}$ (see
Eq.(2.7)). This combination provides the formal solution of the RG
equations (2.2) to the order of interest. If we consider
explicitly the additional condition $T(l)\ll 1$ as $T\to 0$, with
$F^{(0)}_2 (T(l))\simeq {1\over 4} +O\left({1\over T}
e^{-1/T}\right)$, the term $\Lambda _{r}(l)$ which appears in Eq.
(3.18) takes the form :
\begin{equation}
\Lambda_r(l)\simeq 2l-{n+2\over n+8}\ln \left[1+{n+8\over 16}K_{d}
{u_0\over \varepsilon} (e^{\varepsilon l}-1)\right].
\end{equation}         %   (3.19)
In particular for $d=3$, it reduces to:
\begin{equation}
\Lambda_r(l) = 2l-{n+2\over n+8}\ln \left({l\over l_0}+1\right).
\end{equation}         %   (3.20)

\section{Relevant thermodynamic quantities and phase diagram in
the influence domain of the QCP}
 \setcounter{equation}{0}
 \hspace {16pt}
 We now focus on the flow solution (3.18) with $\Lambda_r(l)$ given by
 Eq. (3.19). Since in any case $\Lambda_r(l)>0$, $t_r(l)$
is a relevant parameter. Then, bearing in mind the matching method
[17,18],  we stop the renormalization procedure until a scale
$l=l^*\gg 1$ is reached at which $t_r(l^*)\simeq 1$. This matching
condition, in view of Eq.(3.18),  provides for $l^*$ the
self-consistent equation:
\begin{equation}
e^{-\Lambda_r(l^*)}\approx (r_0-r_{0c}) +{n+2\over 4} K_{d} u_0
T^{2-\varepsilon} I\left({e^{-l^*}\over T},{1\over T}\right)\ .
\end{equation}         %   (4.1)
where
\begin{equation}
I(x,y)= \int_x^ydx'{{x'}^{1-\varepsilon}\over
e^{x'}-1}\left[1+{n+8\over 16}K_{d} u_0 \ln\left({1\over
Tx'}\right)\right]^{-1}\ .
\end{equation}         %   (4.2)
Eqs. (4.1)-(4.2) are valid for $|\varepsilon|\ll 1$ and the case
$d=3$ can be simply obtained taking the limit as $\varepsilon\to
0$.

For a solution $l^*\gg 1$ in the low-temperature limit, Eqs.(3.19)
and (3.20) give :
\begin{equation}
\Lambda_r(l^*)\simeq \left\{
\begin{array}{l}
\lambda_rl^*\qquad ,\qquad \varepsilon\neq 0\\
2l^*\left[1-{1\over 2l^*}\ln \left({l^*\over
l_0}+1\right)^{n+2\over n+8}\right]\quad ,\quad \varepsilon=0
\end{array}\right.
\end{equation}         %   (4.3)
where (see eq. (2.6)) $\lambda_r=2-{n+2\over n+8}\varepsilon$ (for
$\varepsilon>0$) and $\lambda_{r}=2$ (for $\varepsilon<0$).

At this stage, for determining $l^*$, it is convenient to consider
separately the cases $\varepsilon\neq 0$ and $\varepsilon=0$.

\medskip

{(i) \large $\varepsilon\neq 0$.}

For $l^*\gg 1$ and $T(l^*)\ll 1$, Eq. (4.1) can be solved by
iteration yielding the solution:
\begin{equation}
e^{l^*}\simeq \left\{(r_0-r_{0c})+{n+2\over 4}K_{d} u_0 I(c,\infty)
T^{2-\varepsilon}\right\}^{-1 / \lambda_r},
\end{equation}         %   (4.4)
where
\begin{equation}
I(c,\infty)=\int_c^\infty dx{x\over e^x-1} +O(u_0)\simeq
I(0,\infty)+O(u_0^{1\over 2})
\end{equation}         %   (4.5)
with $c=\left({n+2\over 4}K_{d}u_{0}\right)^{1/2}$ and
$I(0,\infty)={\pi^2\over 16}$.

Then, to the order of interest, we can write:
\begin{equation}
e^{l^*}\simeq \left[t_{r_0}(T)\right]^{-1/ \lambda_r},
\end{equation}         %   (4.6)
where
\begin{equation}
t_{r_0}(T)= (r_0-r_{0c}) +{n+2\over 64} \pi^2 K_{d} u_0
T^{2-\varepsilon} \
\end{equation}         %   (4.7)
measures the low-temperature deviation from the QCP.

\medskip

{(ii) \large $\varepsilon = 0$ $(d=3)$.}

In this case, taking into account Eq. (4.3), the self-consistent
equation for $l^*\gg 1$ is:
\begin{equation}
e^{-l^*}\simeq\left\{(r_0-r_{0c}) +{n+2\over 8\pi^2}u_0 T^2 I
\left({e^{-l^*}\over T}, {1\over T}\right)\right\}^{1\over
2}\left({l^*\over l_0}+1\right)^{-{n+2\over 2(n+8)}}\ .
\end{equation}         %   (4.8)
Then, the appropriate low-$T$ solution can be written in the form:
\begin{equation}
e^{l^*}\simeq\left({1\over 2l_0}\right)^{n+2\over
2(n+8)}t_{r_0}^{-{1\over 2}} (T)\left[\ln t_{r_0}^{-1}
(T)\right]^{n+2\over 2(n+8)}
\end{equation}         %   (4.9)
where $t_{r_o}(T)$ is given by Eq. (4.7) simply setting
$\varepsilon=0$:
\begin{equation}
t_{r_0}(T) = (r_0-r_{0c}) +{n+2\over 128} u_0 T^2\ .
\end{equation}         %   (4.10)

With the explicit expressions (4.6) $(\varepsilon\neq 0)$ and
(4.9) $(\varepsilon=0)$ for $l^*$, we have the basic ingredients
for exploring the quantum critical properties in the influence
domain of the QCP.  We can indeed, according to the general RG
philosophy, utilize the scaling relations
for the correlation length $\xi$, the susceptibility $\chi$ and
the singular part of the free energy density. Here we focus on
$\xi$ and $\chi$ for which these relation are:
\begin{equation}
\xi(r_0,u_0,T)\simeq \xi_0e^{l^*}
\end{equation}         %   (4.11)
and (with $\eta=0$):
\begin{equation}
\chi(r_0,u_0,T)\simeq \chi_0 e^{2l^*} \propto \xi^2
\end{equation}         %   (4.12)
where $\xi_0$ and $\chi_0$ are constants inessential for our
present purposes. Then, very near the QCP we have:
\begin{equation}
\xi(r_0, u_0,T)\simeq\xi_0\times \left\{
\begin{array}{lcl}
[t_{r_o}(T)]^{-1/ \lambda_r} & , & \varepsilon\neq 0\\
\left({1\over 2l_0}\right)^{n+2\over 2(n+8)}t_{r_0}^{-{1\over
2}}(T) \left[\ln t_{r_o}^{-1} (T)\right]^{n+2\over 2(n+8)} & , &
\varepsilon=0
\end{array}\right.
\end{equation}         %   (4.13)
and
\begin{equation}
\chi(r_0, u_0,T)\simeq\chi_0\times \left\{
\begin{array}{lcl}
[t_{r_o}(T)]^{-2/ \lambda_r} & , & \varepsilon\neq 0\\
\left({1\over 2l_0}\right)^{n+2\over n+8}t_{r_0}^{-1}(T) \left[\ln
t_{r_0}^{-1} (T)\right]^{n+2\over n+8} & , & \varepsilon=0
\end{array}\right.
\end{equation}         %   (4.14)
where $t_{r_0}(T)$ is given by Eqs. (4.7) and (4.10) for $d\neq 3$
and $d=3$ respectively.

The low-temperature critical line in the ($r_0,T$)-plane, merging in
the QCP, is obtained setting $t_{r_0}(T)=0$ in Eq. (4.7). We have:
\begin{equation}
r_{0c}(T)= r_{0c}-{n+2\over 64}\pi^2 K_{d} u_0 T^{2-\varepsilon}
\end{equation}         %   (4.15)
or, equivalently, for $r_0\le r_{0c}$:
\begin{equation}
T_c(r_0)=\left({64\over \pi^2(n+2)u_0}\right)^{1\over 2}
(r_{0c}-r_0)^{{1\over 2}\left(1+{\varepsilon\over 2}\right)} ,
\end{equation}         %   (4.16)
which are valid also when $\varepsilon=0$.

In terms of $r_{0c}(T)$ or $T_{c}(r_0)$, the critical line deviation
parameter $t_{r_0}(T)$ can be written as:
\begin{equation}
t_{r_0} (T)=r_0-r_{0c}(T)
\end{equation}         %   (4.17)
or
\begin{equation}
t_{r_0}(T) ={n+2\over 64}\pi^2 K_{d} u_0
\left(T^{2-\varepsilon}-T_c^{2-\varepsilon}(r_0)\right).
\end{equation}% (4.18)
In particular this last representation yields:
\begin{equation}
t_{r_0}(T)\approx \left\{
\begin{array}{lclcl}
\displaystyle{n+2\over 32}\pi^2 K_{d} u_0
T_c(r_0)\left[T-T_c(r_0)\right] &,&
\displaystyle r_0\neq r_{0c} &,&\displaystyle \left(T_c(r_0)\neq 0\right) \\
\displaystyle {n+2\over 64}\pi^2 K_{d}u_0 T^{2-\varepsilon}
&,&\displaystyle r_0 = r_{0c} &,&\displaystyle \left(T_c(r_{0c})=
0\right).
\end{array}\right.
\end{equation}         %   (4.19)

From the previous expressions, it is evident that different ways
of approaching the critical line are possible. This is related to
realistic experiments where distinct thermodynamic paths
approaching a critical point may exist. Each of them is
characterized by a specific set of critical exponents and
different sets may be connected by appropriate relations (for
instance a Fisher's renormalization [19] when thermodynamic
constraints are involved). This is just the case of quantum
systems for which several paths of approaching to a critical line
are possible [5]. For the transverse Ising-like models under
study, we will consider the most relevant paths within the
disordered phase ($t_{r_0}(T)>0$) in the ($r_0,T$)-plane:
$L_{r_0}\equiv (r_0\to r_{0c}^+,T=0)$, $L_{T}\equiv (r_0=
r_{0c},T\to 0)$, $L_{r_0-r_{0c}(T)}\equiv (r_0\to r_{0c}^+(T),T\
$fixed$)$ and $L_{T-T_c(r_0)}\equiv (r_0\ $ fixed, $T\to
T_c^+(r_0))$ with $T_c (r_0)\to 0$ as $r_0 \to r_{0c}$.
Correspondently,  for a generic macroscopic quantity $X$ we have
the set of critical exponents $\{x_r\}$ defined by
\begin{equation}
X\sim(r_0-r_{0c})^{-x_r}\quad \hbox{or} \quad X\sim(r_0-r_{0c}(T))^{-x_r}
\end{equation}%(4.20)
along $L_{r_0}$ or $L_{r_o-r_{0c}(T)}$ and the set of critical
exponents  $\{x_T\}$ defined by

\begin{equation}
X\sim T^{-x_T}\quad \hbox {or} \quad X\sim(T-T_c(r_0))^{-x_T}
 \end{equation}%(4.21)
 along  $L_T$ or $L_{T-T_c(r_0)}$.
For instance, for quantum displacive systems [9,10] with
$-r_0=S>0$ and QCP-coordinates ($S=S_c, T=0$) the thermodynamic
paths defined above will be denoted by $L_S$, $L_{S-S_{c}(T)}$,
$L_T$ and $L_{T-T_c(S)}$.

The different paths of interest for us are schematically shown in
Fig.1 for a generic transverse Ising-like system. For sake of
brevity, in the following  we will often speak  of $L_k$-quantum
criticality, with $k=r_0$, $r_0-r_{0c}(T)$, $T$, $T-T_c(r_0)$.

\section{Low-temperature quantum critical properties.}
 \setcounter{equation}{0}
\hspace {16pt} The previous arguments provide all the ingredients
to derive the near-QCP properties around and at $d=3$ for quantum
systems with transverse Ising-like intrinsic dynamics. Here we
focus on the quantum critical behaviour of the correlation length
and the susceptibility along the previously defined thermodynamic
paths. The near-QCP properties of the other thermodinamic
quantities follow immediately from the RG picture. In the next
section we will describe the related crossover phenomena of
experimental interest.

\subsection{$L_{r_0}$-quantum criticality}

\hspace {16pt}
In such a case, from Eqs. (4.7), (4.13) and (4.14)
we have for $d\neq 3$ (consistently with the ($T=0$)-results of
section 2.):
\begin{equation}
\xi\sim \left(r_0-r_{0c}\right)^{-\nu_r}\qquad ,\qquad \chi\sim
\left(r_0-r_{0c}\right)^{-\gamma_r}
\end{equation}         %   (5.1)
where $\nu_r$ and $\gamma_r$ are given by Eqs. (2.8), (2.9) for
$d<3$ and $d>3$, respectively.
For $d=3$, the same equations provide

% equazione (5.2)
\begin{equation}
    \xi \sim (r_{0}-r_{0c})^{-1/2}\left[ \ln
    (r_{0}-r_{0c})^{-1} \right]^{n+2\over 2(n+8)}
\end{equation}

% equazione (5.3)
\begin{equation}
        \chi \sim (r_{0}-r_{0c})^{-1} \left[ \ln
    (r_{0}-r_{0c})^{-1}\right]^{n+2\over n+8} .
\end{equation}
These results are well known, but their independent and unified
derivation signals the full consistency of our present
calculations.

\subsection{$L_T$ - quantum criticality}

\hspace {16pt} Here, the approach to QCP is controlled only by the
temperature for $r_{0}$ fixed to its ($T=0$)-critical value. This
situation is recurrent in several, now accessible, experiments.

From the general Eqs. (4.7), (4.10), (4.13) and (4.14), we
have, for $d\neq 3$
% equazione (5.4)
\begin{equation}
    \xi(T) \sim T^{-\nu_{T}} \quad , \quad \chi (T) \sim T^{-\gamma_{T}}
\end{equation}
with

% equazione (5.5)
\begin{equation}
\nu_{T}=1-{3\over n+8}\varepsilon \qquad ;\qquad \gamma_{T} =
2-{6\over n+8}\varepsilon \qquad ,\qquad \varepsilon > 0
\end{equation}
and
% equazione (5.6)
\begin{equation}
 \nu_{T}=1-{\varepsilon \over 2}\qquad ;
\qquad\gamma_{T}=2-\varepsilon \qquad , \qquad \varepsilon < 0
\end{equation}
For the marginality case $d=3$, where logarithmic corrections to MF
results are expected, we have from the same equations:
% equazione (5.7)
\begin{equation}
    \xi(T) \sim T^{-1}[\ln T^{-1}]^{n+2\over 2(n+8)} \quad \sim \quad
    T^{-1}|\ln T^{2}|^{n+2\over 2(n+8)}
\end{equation}
and
% equazione (5.8)
\begin{equation}
    \chi(T) \sim T^{-2}[\ln T^{-1}]^{n+2\over n+8} \quad \sim \quad
    T^{-2}|\ln T^{2}|^{n+2\over n+8}.
\end{equation}
The previous results for $d<3$ are in full agreement with the
large-n limit predictions [9].

In the  most interesting situation $d=3$, Eqs. (5.7)-(5.8) for
one-component systems yield:

% equazione (5.9)
\begin{equation}
    \xi(T) \sim T^{-1}|\ln T^{2}|^{1\over 6}
\end{equation}
and
% equazione (5.10)
\begin{equation}
    \chi(T) \sim T^{-2}|\ln T^{2}|^{1\over 3} ,
\end{equation}
which are again in full agreement with the results
obtained by Schmeltzer [20-23] via a direct
($d=3$)-field theoretic approach, and by Caramico {\it et al}. [5]
using the temperature dependent linear scaling field method.

As concerning the $T$-driven critical behaviours for $\varepsilon
< 0$ ($d>3$), it must be noted that the $L_{T}$-critical exponents
correspond to the known $L_{r_{0}}$-Gaussian ones which give the
leading corrections [23] to the MF behaviours. In particular, the
exponents $\nu_{T}$, $\gamma_{T}$, given by Eqs. (5.6),
characterize the corrections to the unknown dominant singularities
for $d>3$, corresponding to the MF ones along $L_{r_{0}}$. The
explicit RG calculation of the exponents along $L_{T}$, involving
a study of the role played by irrelevant dangerous variables, is
beyond the purpose of the present paper.

\subsection{$L_{r_0-r_{0c(T)}}$ - quantum criticality}

\hspace {16pt}
The starting points are always the  unified Eqs.
(4.13), (4.14) with the representation (4.18) for $t_{r_{0}}(T)$.
Then, it immediately follows:

(i) for $d\neq 3$:
% equazione (5.11)
\begin{equation}
    \xi \sim (r_{0}-r_{0c}(T))^{-\nu_{r}}, \chi \sim
    \left( r_{0}-r_{0c}(T)\right)^{-\gamma_{r}}
\end{equation}
with the same $L_{r_{0}}$ - critical exponents (2.8)-(2.9);

(ii) for $d= 3$:

% equazione (5.12)
\begin{equation}
    \xi \sim (r_{0}-r_{0c}(T))^{-{1\over2}}\left[
    \ln (r_{0}-r_{0c}(T))^{-1}\right]^{n+2\over 2(n+8)}
\end{equation}
and
% equazione (5.13)
\begin{equation}
    \chi \sim (r_{0}-r_{0c}(T))^{-1}\left[
    \ln (r_{0}-r_{0c}(T))^{-1}\right]^{n+2\over (n+8)}.
\end{equation}

These low-$T$ results in the influence domain of QCP suggest new
possible experiments for realistic systems, driving the transition
with a nonthermal parameter (transverse field, elastic strenght
parameter, pressure, etc.) at fixed $T\neq 0$.

\subsection{$L_{T-T_{c}(r_0)}$ - quantum criticality}

\hspace {16pt}Using the representation (4.18) for $t_{r_{0}}(T)$,
in the case $\varepsilon >0 \quad (d<3)$ Eqs. (4.13), (4.14)
yield:
% equazione (5.14)
\begin{equation}
    \xi(T) \sim \left(T^{2-\varepsilon}-T_{c}^{2-\varepsilon}(r_{0})
    \right)^{-{1\over\lambda_{r}}}
\end{equation}
and
% equazione (5.15)
\begin{equation}
    \chi(T) \sim \left(T^{2-\varepsilon}-T_{c}^{2-\varepsilon}(r_{0})
    \right)^{-{2\over\lambda_{r}}}.
\end{equation}
For $d>3$ a quite similar scenario occurs and we do not
consider this case explicitly.

From  Eqs. (5.14), (5.15) two different behaviours governed by
the temperature occur according to $r_{0}\neq r_{0c}\quad
(T_{c}(r_{0})\neq 0)$ or $r_{0}=r_{0c}\quad  (T_{c}(r_{0c})=0)$.
In the last case Eqs. (5.14) and (5.15) reduce to Eqs. (5.4),
(5.5) characterizing  the behavious of $\xi$ and $\chi$ along the
line $L_{T}$. The situation is quite different when $r_{0}\neq
r_{0c}$. In this case $t_{r_0}(T)$ $\propto T-T_{c}(r_{0})$  and
Eqs. (5.14), (5.15) reduce to:
% equazione (5.16)
\begin{equation}
    \xi(T) \sim \left( T-T_{c}(r_{0})
    \right)^{-{1/\lambda_{r}}}
\end{equation}
and
% equazione (5.17)
\begin{equation}
    \chi(T) \sim \left( T-T_{c}(r_{0})
    \right)^{-{2/\lambda_{r}}}
\end{equation}
which imply
% equazione (5.18)
\begin{equation}
    \nu_{T}\equiv \nu_{r}={1\over2}(1+{n+2\over 2(n+8)}\varepsilon)
\end{equation}
% equazione (5.19)
\begin{equation}
    \gamma_{T}\equiv \gamma_{r}=1+{n+2\over 2(n+8)}\varepsilon \quad ,
\end{equation}
respectively.
Thus, for $r_{0}\rightarrow r_{0c}$, a crossover phenomenon takes
place which is described by the general equations (5.14)-(5.15)
(See Sect. 6 below).

For $d=3$, a Gaussian-like behaviour with
logarithmic corrections occurs as expected for a marginality case.
Eqs. (4.13), (4.14) imply indeed:
% equazione (5.20)
\begin{equation}
    \xi(T) \sim (T^{2}-T_{c}^{2}(r_{0}))^{-{1\over2}}\left[
    \ln (T^{2}-T_{c}^{2}(r_{0}))^{-1}\right]^{n+2\over 2(n+8)}
\end{equation}
and
% equazione (5.21)
\begin{equation}
    \chi(T) \sim (T^{2}-T_{c}^{2}(r_{0}))^{-1}\left[
    \ln (T^{2}-T_{c}^{2}(r_{0}))^{-1}\right]^{n+2\over (n+8)}.
\end{equation}
These reduce to Eqs. (5.7), (5.8) when $r_{0}=r_{0c}$
 and to:
% equazione (5.22)
\begin{equation}
    \xi(T) \sim (T-T_{c}(r_{0}))^{-{1\over2}}\left[
    \ln (T-T_{c}(r_{0}))^{-1}\right]^{n+2\over 2(n+8)}
\end{equation}
and
% equazione (5.23)
\begin{equation}
    \chi(T) \sim (T-T_{c}(r_{0}))^{-{1}}\left[
    \ln (T-T_{c}(r_{0}))^{-1}\right]^{n+2\over (n+8)}
\end{equation}
when $r_{0}\neq r_{0c}$. As we
see, also for $d=3$ a crossover phenomenon occurs for
$r_{0}\rightarrow r_{0c}$ which is described by eqs. (5.20), (5.21) but
now it involves logarithmic corrections to the power law.

For $n=1$, the general Eqs. (5.20), (5.21) reduce to:
% equazione (5.24)
\begin{equation}
    \xi(T) \sim (T^{2}-T_{c}^{2}(r_{0}))^{-{1\over2}}\left|
    \ln (T^{2}-T_{c}^{2}(r_{0}))\right|^{1\over 6}
\end{equation}
and
% equazione (5.25)
\begin{equation}
    \chi(T) \sim (T^{2}-T_{c}^{2}(r_{0}))^{-{1}}\left|
    \ln (T^{2}-T_{c}^{2}(r_{0}))\right|^{1\over 3}
\end{equation}
again in agreement with the corresponding field-theoretic results
by Schmeltzer [20-22].

The near-QCP behaviours of other thermodynamic quantities can be
similarly extracted from the rescaling relations of the singular
part $F_{s}$ of the free energy density with the help of the above
expressions of the deviation parameter $t_{r_{0}}(T)$. This lies
on the fact that, by iteration of the RG transformation until the
scale $l^{*}$ is reached, one can write $F_{s}\sim e^{(d+1)l^{*}}$
where $e^{l^{*}}$ is given by Eqs. (4.6) - (4.10) and (4.18). For
some explicit results we limit ourselves to the thermodynamic
paths $L_{T}$ and $L_{T-T_{c}(r_{0})}$ for $d\leq 3$, for which
the most interesting features occur.

For $\epsilon > 0$, by approacching the QCP along $L_{T}$ one
finds

% equazione 5.26
\begin{equation}
     F_{s}(T) \sim T^{(d+1)\nu_{T}}, \hspace{.5 cm} {C_{s}(T)\over T}\sim
     T^{-\alpha_{T}}
\end{equation}
with
% equazione 5.27
\begin{equation}
  \alpha_{T}=2-(d+1)\nu_T = -2+{n+20 \over n+8}\epsilon \hspace{.5 cm}.
\end{equation}
In Eq. (5.26), $C_{s}(T) = -T{\partial^{2}F_{s}\over
\partial T^{2}}$ is the contribution to the specific heat arising
from the singular part of the free energy density [28]. As $T
\rightarrow T_{c}^{+}(r_{0})$ along $L_{T-T_{c}(r_{0})}$, Eq.
(4.18) yields:

% equazione 5.28
\begin{equation}
     F_{s}(T)\sim
(T^{2-\epsilon}-T_{c}^{2-\epsilon})^{(d+1)\nu_{r}}\hspace{.5 cm}.
\end{equation}
Of course, from this relation, Eqs. (5.26) are reproduced as
$r_{0}\rightarrow r_{0c}$ while, for $r_{0}\neq r_{0c}$
$(T_{c}(r_{0})\neq 0)$, one obtains $F_{s}\sim
(T-T_{c}(r_{0}))^{(d+1)\nu_{r}}$ and hence $C_{s}(T)/
T_{c}(r_{0})\sim C_{s}(T)\sim (T-T_{c}(r_{0}))^{-\alpha_{T}}$ with
$\alpha_{T}\equiv \alpha_{r}={4-n\over 2(n+8)}\epsilon$ . Here
$\alpha_{r}$ is the critical exponent which characterizes the
$(T=0)$-behaviour of the specific heat-like quantity
$C_{r}=-{\partial^{2}F_{s}\over \partial r_{0}^{2}}$ as
$r_{0}\rightarrow r_{0c}^{+}$ [5]. Thus, also for $C_{s}(T)/ T$ a
crossover occurs as $r_{0}\rightarrow r_{0c}$ (or
$T_{c}(r_{0})\rightarrow 0$)[28].

It is worth noting that, in any case, the $x_{T}$-exponents
$\alpha_{T}$ and $\nu_{T}$, as $\alpha_{r}$ and $\nu_{r}$, satisfy
the $T$-hyperscaling relation $2-\alpha_{T}=(d+1)\nu_{T}$.

At $d=3$, as for $\chi$ and $\xi$, logarithmic corrections occur.
Indeed, from Eqs. (4.9)-(4.10) we have
% equazione 5.29
\begin{equation}
     F_{s}(T)\sim T^{4}|\ln T^{2}|^{-{2(n+2)\over n+8}}
     \hspace{.2 cm},\hspace{.3 cm}{C_{s}(T)\over T}\sim T^{2}|\ln
     T^{2}|^{-{2(n+2)\over n+8}}\hspace{.2 cm} ,
 \end{equation}
as $T\rightarrow 0$ along $L_{T}$. In contrast, if we approach the
critical line at fixed $r_{0}\neq r_{0c}$ and $T\rightarrow
T_{c}^{+}(r_{0})$ we have,
% equazione 5.30
\begin{equation}
     F_{s}(T)\sim (T^{2}-T_{c}^{2}(r_{0}))^{2}|\ln
(T^{2}-T_{c}^{2}(r_{0}))|^{-{2(n+2)\over n+8}}
     \sim (T-T_{c}(r_{0}))^{2}|\ln (T-T_{c}(r_{0}))|^{-{2(n+2)\over n+8}}
     \end{equation}
and hence a mean-field behaviour in $(T-T_{C}(r_{0}))$, with
logarithmic corrections for ${C_{s}(T)\over T}$, takes place.
Then, at $d=3$, a crossover occurs also for ${C_{s}(T)/ T}$ as
$r_{0}\rightarrow r_{0c}$ but now logarithmic corrections are
involved.

\section{Crossover phenomena and comparison with experiments}
\setcounter{equation}{0}

 \hspace {16pt}In the previous subsection 5.4, we have
shown that a crossover between the $L_{T-T_{c}(r_{0})}$-critical
regime and $L_{T}$-one occurs as $r_{0}\rightarrow r_{0c}$.

For $\varepsilon \neq 0$, when logarithmic corrections are not
involved, this crossover phenomenon can be conveniently described
in terms of ``effective exponents". Here we focus on
susceptibility and determine the appropriate effective exponents
$\gamma^{eff}_{T}=2\nu^{eff}_{T}$, where $\nu^{eff}_{T}$ denotes
the effective exponent for the correlation length. Other effective
exponents (as $\alpha_T ^{eff}$, see below) can be derived
similarly.

We start from the general asymptotic relation (5.15) with
$2/\lambda_{r}=2\nu_{r}=\gamma_{r}$ given by Eqs.(2.8)-(2.9). As
clarified before, this relation, as $T \rightarrow
T_{c}^{+}(r_{0})$ from the disordered phase $(t_{r_{0}}(T)>0)$,
contains both the behaviours when $T_{c}(r_{0c})=0 \quad
(\gamma_{T} = (2-\varepsilon)\gamma_{r})$ and $T_{c}(r_{0})\neq 0
\quad (\gamma_{T}=\gamma_{r})$. Then, Eq. (5.15) allows us to
extract an effective exponents $\gamma^{eff}_{T}$ such that
% equazione (6.1)
\begin{equation}
    \gamma_{r} \leq \gamma_{T}^{eff} \leq (2-\varepsilon)\gamma_{r}.
\end{equation}
This inequality, as $d\rightarrow 3$, is in full agreement with
experimental data for quantum ferroelectric oxides [24] and other
displacive systems [25]-[27]
except for logarithmic corrections which can be taken into account
from relations (5.22)-(5-25).

To prove the inequality (6.1), we
first observe that, from the conventional definition of
``effective critical exponent" for a generic quantity $\chi\sim
t^{-\gamma_{eff}}$
% equazione (6.2)
\begin{equation}
    \gamma^{eff}=-{\partial \ln \chi \over \partial \ln t} ,
\end{equation}
 Eq. (5.15) leads to
 % equazione (6.3)
\begin{equation}
    \gamma^{eff}_{T}(\tau)= (2-\varepsilon)\gamma_{r} {1-\tau \over
    1-\tau^{2-\varepsilon}}
\end{equation}
where $\tau = T_{c}(r_{0})/T$  ($0\leq \tau\leq 1$)
 is the appropriate crossover
parameter for the problem. Then, since
$1-\tau^{2-\varepsilon}=(2-\varepsilon)(1-\tau)$ as $\tau
\rightarrow 1$, it is easy to check that
the effective exponent (6.3) satisfies the inequality (6.1), with
$\gamma^{eff}_{T}(0)=(2-\varepsilon)\gamma_{r}$ and
$\gamma^{eff}_{T}(1)=\gamma_{r}$ .

With $d\rightarrow 3^{\mp}$, Eq. (6.3) becomes
 % equazione (6.4)
\begin{equation}
    \gamma_{T}^{eff}(\tau)={2\over 1+\tau}
\end{equation}
satisfying the inequality $1\leq \gamma_{T}^{eff}(\tau)\leq 2$.

A similar analysis can be performed to determine the effective
exponent $\alpha_{T}^{eff}$ for ${C_{s}(T)/ T}$. We find

% equazione (6.5)
\begin{equation}
     \alpha_{T}^{eff}(\tau)=2-(4-\epsilon)\nu_{r}f_{\epsilon}(\tau)
     \end{equation}
which crossovers between $\alpha_{T}^{eff}(0)=\alpha_{T}$ (Eq.
5.27) and $\alpha_{T}^{eff}(\tau =1)=\alpha_{r}$.

For $d=3$, an analogous crossover between two marginal regimes
(involving logarithmic corrections) occurs as $r_{0}\rightarrow
r_{0c}$. This crossover, as already mentioned before, can be
described by means of the general relations (5.20)-(5.21) and the
corresponding one for $F_s(T)$ but now effective exponents can not
be exactly defined as for $d\neq 3$ due to the presence of
logaritmic corrections which are experimentally inaccessible.

All the previous RG predictions concerning the $L_{k}$-quantum
criticality and the crossover phenomena around and at $d=3$ are
quite consistent with experiments on transverse Ising-like quantum
systems [8]  exhibiting a quantum phase transition. Relevant
examples are the quantum displacive systems with quantum
structural phase transitions [24]-[27] and, in particular,
ferroelectric oxides as $K_{1-x}Na_{x}TaO_{3}$
 and $KTa_{1-x}Nb_{x}O_{3}$  with fixed
composition $x$ [24]. The experimental results for this class of
systems, relevant for a direct comparison with our RG theoretical
results, can be summarized as follows (see Fig.2  for a schematic
picture of a typical phase diagrams and for related notations).

1. Approaching the QCP ($S_{c}, T=0$) along $L_{S}$, the
susceptibility varies with the interaction parameter $S$
(proportional to the pressure) as :
 % equazione (6.6)
\begin{equation}
    \chi \sim (S_{c}-S)^{-\gamma_{S}}
\end{equation}
with $\gamma_{S}=1$. This agrees with our RG results (5.1)-(5.3);

2. Along $L_{T}$, at the quantum displacive limit ($S=S_{c}$),
$\chi$ varies with the temperature as:
 % equazione (6.7)
\begin{equation}
    \chi(T) \sim T^{-\gamma_{T}}
\end{equation}
with $\gamma_{T}=2$, in agreement with the theoretical results
(5.4)-(5.10);

3. In the low-temperature limit, the transition temperature
$T_{c}(S)$ as a function of $S$ (or pressure) is expressed as:
 % equazione (6.8)
\begin{equation}
    T_{c}(S)\sim (S-S_{c})^{1\over2}
\end{equation}
which is exactly reproduced by Eq. (4.16) with $\varepsilon =0$;

4. Approaching the critical line along $L_{T-T_{c}(S)}$, one has:
 % equazione (6.9)
\begin{equation}
    \chi\sim (T-T_{c}(S))^{-\gamma_{T}}
\end{equation}
where $\gamma_{T}$ increases from $\gamma_{T}=1$, when
$T_{c}(S)\neq 0$, to the value $\gamma_{T}=2$ when
$T_{c}(S)\rightarrow 0$ as $S\rightarrow S_{c}$ (see Fig. 5 by
Samara [24] for $K_{1-x}Na_{x}TaO_{3}$). This signals a crossover
by continuous variation of $T_{c}(S)$ towards zero when $S$
decreases to its quantum displacive limit $S_{c}$, in full
agrement with our theoretical predictions (6.1)-(6.4). Of course,
logarithmic corrections, expected in the marginal situation $d=3$,
are experimentally undetectable but they have to emerge from a
consistent theory as it happens in the present RG calculations.
\newpage

\section{Concluding remarks}

\hspace {16pt}For a big amount of microscopic systems exhibiting a
continuous QPT, the $T$-dependent RG equations are known since
more than twenty years [13], a part from few next relevant
modifications in a number of interesting cases [29,30].

Recent experiments,
involving compounds for which unusual low-$T$ properties are
observed, have motived the new tendency [1]-[6] to solve, with a minimum of reliable approximations,
the very complicate RG equations with the aim to
extract a lot of macroscopic information to be compared with
available experimental data.

In the present paper, following this very
fruitful and promising tendency, we have
derived explicit low-$T$ expressions of relevant
thermodynamic quantities for a wide class of quantum systems
solving the related one-loop RG equations in the low-$T$ regime.
Our aim was to explore the quantum critical
behaviours within the domain of influence of a QCP and to compare our
results with experiments, where different intriguing ways of approaching
the QCP may occur. We have focused on transverse Ising-like models
which are the basic starting point for exploring theoretically
the thermodynamic properties of several realistic systems of
experimental interest which exhibit a QPT. It is a fortunate
feature that, for this class of models, the RG works just around
and at $d=3$. This allowed us to obtain practically exact low-$T$
results, so that a direct comparison with experiment data becomes
suitable.

Our low-$T$ RG predictions have been shown to be in remarkable
agreement with the experimental data for quantum displacive
systems, as the ferroelectric oxides, for which accurate
experiments have been performed around their QCP's, where the
quantum fluctuations are expected to strongly influence the
response of the systems.

Finally, from a methodological point of view, we wish to underline
that our RG analysis, based on a $T$-dependent solution of the
flow equations which avoids ``ad hoc" assumptions, provides a
valid support to the so called ``temperature-dependent linear
scaling fields method"[5], formulated beyond the traditional
schemes of the RG approaches.

\newpage

FIGURE CAPTIONS

Fig.1. Schematic low-temperature phase diagram for a generic
transverse-Ising-like model with the thermodynamic paths
$L_{r_0}$, $L_T$, $L_{{r_0}-r_{0c}(T)}$ and $L_{T-T_c(r_0)}$
approaching the QCP.

FIG.2.  Schematic low-temperature phase diagram for quantum
displacive systems with the thermodynamic paths $L_{S}$, $L_{T}$
and $L_{T-T_{c}(S)}$ within the disordered phase in the influence
domain of the QCP.
\newpage

\centerline{\Large \bf References.} \vspace{2cm}
\begin{enumerate}
\item S. Sachdev,  Quantum Phase Transitions, Cambridge University
Press, Cambridge, (1999).
\item U. Z\"{u}licke and A.J. Millis, Phys. Rev. B51 (1995) 8996.
\item A. J. Millis, Phys. Rev. B 48 (1993) 7183.
\item S. Sachdev, T. Sentil, and R. Shankar, Phys. Rev. B 50(1994)
258.
\item A. Caramico D'Auria, L. De Cesare and I. Rabuffo, Physica A 243 (1997)
152.
\item M. Crisan, D. Bodea, J. Grosu and I. Tifrea, J. Phys. A: Math. Gen. 35 (2002)
239.
\item M. Crisan, D. Bodea and J. Grosu, cond-mat/0207712.
\item B.K. Chakrabarti, A. Dutta and P. Sen, Quantum Ising Phases
and Transitions in Transverse Ising Models, Springer, Berlin,
1996.
\item R. Morf, T. Schneider, and E. Stoll, Phys. Rev. B 16 (1977)
462.
\item R. Oppermann and H. Thomas, Z. Physik B 22 (1975) 387.
\item L. De Cesare, Rev. Solid State Sci. 3 (1989) 71.
\item A. Caramico D'Auria, L. De Cesare and I. Rabuffo, Phys. Lett.
A, in press.
\item J. Hertz, Phys. Rev. B 14 (1976) 1165.
\item S. Ma, Modern Theory of critical phenomena, W.A. Benjamin
Inc., London, 1976.
\item P. Pfeuty and G. Toulouse, Introduction to the
Renormalization Group and Critical Phenomena, Wiley, London
(1977).
\item N. Goldenfeld, Lectures on Phase transitions and the
Renormalization Group,  Addison-Wesley Pub. co., New York, 1993.
\item J. Rudnick and D.R. Nelson, Phys. Rev. B 13 (1976) 2208.
\item H. -O. Hener and D. Wagner, Phys. Rev. B  40 (1989) 2502.
\item M. E. Fisher, Phys. Rev. 176 (1968) 257.
\item D. Schmeltzer, Phys. Rev. B 28 (1983) 459.
\item D. Schmeltzer, Phys. Rev. 29 (1984) 2815.
\item D. Schmeltzer, Phys. Rev. 32 (1985) 7512.
\item E.H. Auge and J.S. H\"{o}ye, Physica Scripta T 44 (1992) 42.
\item G. A. Samara, Physica B 150 (1988) 179.
\item U. T. Hochli  and L. A. Boatner, Phys. Rev. B 20 (1979) 266.
\item H. Bilz, Phys. Rev. B 22 (1980) 359.
\item K.A. Muller, J. Japan  Appl. Phys. 24 (Suppl. 2) (1985) 89.
\item In the quantum regime the exponent for specific heat can be
defined in two ways since, for $T\rightarrow 0$, the additional
$T$-factor in its definition reduces the singularity by one power
in $T$. Then one defines $C_{s}\sim T^{-\tilde{\alpha}_{T}}$ and
${C_{s}(T)\over T}\sim T^{-\alpha_{T}}$, where
$\alpha_{T}=\tilde{\alpha}_{T}+1$ corresponds to the usual
specific heat exponent in the theory of critical phenomena.
\item T.R. Kirkpatric and D. Belitz, Quantum Phase Transition in
Electronic Systems, in Electron Correlation in the Solid State, N.
H. March (editor), Imperial College Press, Oxford (1999).
\item  T. Vojta, Ann. Phys. 9 (2000) 403.
\end{enumerate}
\end{document}